\documentclass[sigconf, nonacm]{acmart}

\AtBeginDocument{%
  \providecommand\BibTeX{{%
    \normalfont B\kern-0.5em{\scshape i\kern-0.25em b}\kern-0.8em\TeX}}}

\setcopyright{acmcopyright}
\copyrightyear{2018}
\acmYear{2018}
\acmDOI{10.1145/1122445.1122456}

\acmConference[Woodstock '18]{Woodstock '18: ACM Symposium on Neural
  Gaze Detection}{June 03--05, 2018}{Woodstock, NY}
\acmBooktitle{Woodstock '18: ACM Symposium on Neural Gaze Detection,
  June 03--05, 2018, Woodstock, NY}
\acmPrice{15.00}
\acmISBN{978-1-4503-XXXX-X/18/06}

\usepackage{graphicx}
\usepackage{subcaption}
\usepackage{multirow}
\usepackage{color,soul}
\usepackage{array, caption, tabularx,  ragged2e,  booktabs}
\usepackage{tabularx}
\usepackage{xcolor}
\usepackage{hyperref}
\usepackage{subcaption}
\usepackage{ulem}
\usepackage{color, colortbl}

\setcopyright{none}
\begin{document}

\title[Creative Compensation]{Creative Compensation (CC): Future of Jobs with Creative Works in 3D Printing}

\author{Chen Liang}
\email{clumich@umich.edu}
\affiliation{
  \institution{University of Michigan}
  \city{Ann Arbor}
  \state{Michigan}
}

\author{Nahyun Kwon}
\email{nahyunkwon@tamu.edu}
\affiliation{
  \institution{Texas A\&M University}
  \city{College Station}
  \state{Texas}
}

\author{Jeeeun Kim}
\email{jeeeun.kim@tamu.edu}
\affiliation{
  \institution{Texas A\&M University}
  \city{College Station}
  \state{Texas}
}

\renewcommand{\shortauthors}{Chen Liang and Jeeeun Kim}

\newcommand{\todo}[1]
 {{\color{orange} \textbf{\textit{\underline{TODO}}}}}
\newcommand{\jk}[1]
{{\fontfamily{cmss}\selectfont \bfseries \color{violet} JK: #1 }}

\begin{abstract}

With the continuous growth of online 3D printing community 
and the democratization of 3D printers, 
growing number of users start sharing their own 3D designs on open platforms, enabling a wide audience to search, download, and 3D print models for free.
Although sharing is mostly for altruistic reasons at first, open platforms had also created potential job opportunities to compensate creative labors.
This paper analyzes new job opportunities emerged in online 3D printing social platforms and patterns of seeking compensations, and reveals various motivations for
posting creative 
content online.
We find that
offering exclusive membership through subscriptions, selling final products
or printing services through web stores, and 
using affiliate links are primary means of earning profits, while there exist gaps between creators' expectations and realities.
We show that various socio-economic promises emerged, leading to a win-win situation for both creators to gain extra income and audiences to have access to more quality content. We also discuss future challenges that need to be addressed, such as ethical use of opensource content.

\end{abstract}

\ccsdesc[500]{Human-centered computing~Human computer interaction (HCI)}

\keywords{Compensation, open contents market, creative designs, 3D printing, content sharing, online communities }


\maketitle

\section{Introduction}
As 3D printers become increasingly accessible to individuals and small organizations in the past decades\cite{RN0}, 3D printing communities also keep expanding.
While designers may share content online because of altruistic reasons or the rise of maker culture,
it has been shown that financial compensation also becomes one of the common motivations to keep creators sharing and exhibiting their creative activities through open platforms \cite{RN4}.
As the community becomes increasingly inclusive, it has been observed that a new online gig economy appears. Various methods of financial transactions occurred among individuals, including providing on-demand printing services, accepting tips, providing paid membership to offer exclusive designs, and selling 3D printed final products. Designers could utilize these methods to make additional profits from 3D printing social platforms.
For example, freelancers could spend more time on creation and can possibly reduce the length of their day time jobs, retirees could started to turn their hobby projects to receive credits, and entrepreneur could launch startup projects on these communities.
Especially with the current economic downturn impacted by the COVID-19 pandemic, which had many people sheltered at home and work remotely, online gig economy can present potential to supplement decreasing income caused by being retreated from their regular jobs \cite{UpWorkReport}. 
In addition, crowdfunding and sharing platforms in general can also contribute to the growth of the 3D printing innovation ecosystem \cite{RN2}.

The goal of this research is to have a closer look at different methods of compensation 
that support creators in 3D printing communities based on their digital contents that are shared for free, and to summarize potential 
new socio-economic and socio-technical impact. 
Although creative compensation becomes increasingly popular in 3D printing communities, the understanding of what are specific type of \textit{jobs} that 3D printing enthusiasts can participate,
what are promising application categories, and what services or technology help creators online get paid remains limited.
Specifically, this study will focus on how implicit job opportunities have been emerged within 3D printing community and 
how participating members are compensated,
what are creators’ motivations and expectations, and what are the realities. Through this work, we provide answers to the following research questions: 

    \textbf{RQ1.} What are new economic opportunities sparked by 3D printing technology and open 3D content sharing platforms, and how do existing services help creative workers get paid?

    \textbf{RQ2.} With what 3D models do creators have more opportunities to get paid, and what are motivations and aims of creators for seeking profits?

    \textbf{RQ3.} What are unique challenges and implications associated with this community and technology support?

To achieve this goal, we take three methodologies, netnography, data analysis, and an online survey. We primarily focus on Thingiverse, which is the biggest 3D printing community\cite{thingiverse2015downloads}, and as we noticed that many 3D contents on Thingiverse implicitly or explicitly solicit compensation
, we collected all designs on Thingiverse and analyzed different methods for making a profit from their shared designs, including direct ways of seeking compensation (e.g., donation, selling products) and indirect ways (e.g., affiliate links). 
In addition, we also conducted a survey on Thingiverse to gain in-depth understanding about creators, how the revenues they received influence their will to continue producing creative works. 
The collected information provides additional insights to understand creative workers in 3D printing community, which helps us discover unique challenges to shape future jobs to bring our attention to the real impact by imagining the future of jobs. 

Our contributions are as the following:
\begin{itemize}
    \item We provide the list of opportunities that creators in online 3D printing community can make money from open contents 
    \item We show the type of contents that 3D printing enthusiasts can jump in to participate 
    \item We reveal creators' motivation, expectation, and the gap with the reality
    \item We discuss emergent issues, concerns, and implications in expanding the opportunities where some of them can be potential jobs
\end{itemize}
\section{Background}

\subsection{Sharing Creative Talents via Social Platforms}
During the recent blooming of the internet culture, creativity has been playing a role in generating and sharing contents.
With the rise of public platforms that engage people's broad participation into Twitter, YouTube, Instagram, Reddit, and many more, increasing numbers of users share ideas, thoughts, troubleshooting tips, design portfolio, icons, and even how-to videos online \cite{cook2009contribution, settles2013let}.  
Since the `Expert Amateur' in DIY communities has been proposed in 2010 \cite{ExpertAmateur}, broadened accessibility to personal fabrication machines and online communities have revealed new opportunities for bidirectional communication via contents.
Participating in the development of free and opensource content seems to be highly voluntary, particularly in opensource software community \cite{lakhani2003hackers}. 
Internet users might be motivated by internal \& external factors, such as the enjoyment of creating new contents, the sense of rewarding that they are helping others, and the social reputation they received from these communities \cite{von2012carrots}. 
Albeit absence of direct compensation, these motivators have undoubtedly enriched open source communities \cite{alexander2002working} including Thingiverse, which recently became the largest online 3D modeling community \cite{thingiverse2015downloads}.
Firstly established to promote the use of desktop 3D printers with fun and exciting 3D models, Thingiverse is now a much more diverse platform that allows people to share assistive devices \cite{SharingisCaring}, artifacts to augment everyday objects \cite{AugmentedFab}, tangible literacy materials for children \cite{TactilePictures}, 3D scanned models of artistic sculpture from museums, and many more.

\subsection{Beyond Free Contents: Contents Markets that Create Financial Transactions}
Thinking beyond open contents market, users start to realize their creative invention that were 
shared and admired by many others can be converted into real products.
Although free content and the collaborative nature of open source culture are main drivers for innovation today \cite{raasch2013innovation, lakhani2003hackers, roberts2006understanding}, it also shifted to platform-oriented self-employment market where people  were often paid for their labors online \cite{drahokoupil2016platform, farrell2016paychecks, kassi2018online}, for example, cooking classes and coding workshops on Fiverr \footnote{https://www.fiverr.com/}, Chegg\footnote{https://www.chegg.com/tutors}, or Upwork\footnote{https://www.upwork.com/}.
While creative workers seemed upset at sharing secret know-hows for free that they thought it will reduce their revenues,
they realized that platforms can serve as a playground to showcase and market their expertise, thus they are granted more opportunities to sell their unique talents by attracting more viewers \cite{lipsononline, kim2017mosaic}.
With growing number of patents that introduce the concept of monetization of online and digital contents \cite{brougher2009monetization, tran2014monetization}, many creators make a living from supporters by accepting tips and donations and bit cheers \cite{TwichEarnings}, contract for projects \cite{kenney2019work}, consulting \cite{wang2017big}, even establishing startups or local studios supported by crowdfunding \cite{macht2014benefits, RN4} (e.g., Kickstarters, IndieGogo).
Established with the opensource culture, Makerbot also set up opensource community Thingiverse to share their know-hows, models, and more, and the prosperity of user-generated contents and the increase of people's interests positively affected sales of their products \cite{west2016complementarity}.
3D printing communities are so special in that (1) it was fostered by democratization of 3D printers with 16 key patents expired in 2013-2014, so more and more people can afford quality printers at home \cite{PrinterPatentsExpire},
but (2) audiences still might not approach those contents directly due to lack of expertise or cannot afford machines, wanting someone to do it for them \cite{AnyoneCanPrint}.
Therefore, 3D printing stores/services also became much popular, empowering everyone to print custom designs that either they self-designed or customized from the free models downloaded online \cite{pryor2014implementing, mai2016customized}. This service can be supported by anyone who owns 3D printers everywhere in the world, and it redefines manufacturing by inviting many home-manufacturers with low-cost printers to participate in the manufacturing industry. 
Our focus here is what unique opportunities and challenges are newly created with abundant and diverse  public 3D contents and open platforms that call for participation from creators, manufacturers, customers, and even distributors. 

\subsection{Technology to Help Individuals Participate in Online Gig Economy}
If we zoom out, there are flourishing platforms that help people gain profit from extra utilities and time by enabling people to easily freelance. Platforms such as Uber, AirBnB, and UpWork have showed the potential of this method \cite{sundararajan2017sharing}. 
In addition to the platforms that help forming gig economy even without expertise, such as Amazon Mechanical Turks for completing very simple tasks \cite{lakhani2004open}, growing numbers of platforms and services have been assisting financial transactions between individuals through emerging markets.
These platforms offer individuals who need more flexibility a chance to be paid. 
Some platforms are specifically designed for supporting creative works, such as Patreon \footnote{https://www.patreon.com/},
which mainly provide membership services.
Through membership platforms, creators and crowd supporters are connected in a direct and continuous manner, such as monthly subscription
\cite{RN4}. 
Although revenue that creators get could be highly skewed (e.g., \$2,500/mon. for top 1\% campaign vs. pennies for the majority),
the monthly subscription mechanism still brings a substantial number of creators a steady revenue stream that presumably big enough to let them focus their time and energy on their creative work\cite{RN4}.
It also gives a chance for creators to improve their work and for supporters to access more quality contents
\cite{RN4}.
In addition to small amount of revenues, Kickstarters and IndieGogo served as platforms to support initiating small business from creative idea.
Researchers have analyzed this specific user-entrepreneur business models in 3D printing, which is defined as the commercialization of a new product and/or service by an individual or group of individuals who are also users of that product and/or service \cite{RN1}. 
They could provide a wide range of services including 3D printing, consulting, and prototyping, to accommodate customers’ diverse needs in 3D printing due to their insufficient design skills, expertise, 
the technology barrier to setup the 3D printer, and more \cite{RN1}.\par

Examining currently existing data will also provide additional insights on creators’ expectation of how projected earnings will potentially influence the creators’ future path, and what are critical challenges to craft the future of job growth especially with 3D printing.







\section{Methods}

\subsection{Netnography: Studying Thingiverse as an Open 3D Content Marketplace}
Netnography is an online research method that follows ethnography, conducted to understand social interaction in modern digital platforms and communications \cite{Netnography}.
With netnography, we were able to focus on reflections from existing data provided by the largest online 3D printing community Thingiverse, with largely inclusive, but not intrusive data. 
First, we manually looked into popular Thingiverse designs that are published as web documents, consisting of textual information (title, tags, categories, description about the design) and images. 
From the initial investigation, we found that many 3D designers are seeking profits by inviting audiences to visit external URLs of their exclusive membership service accounts and web stores, PayPal or Bitcoin accounts to donate tips, Shapeways stores to get paid for design, and more. 
Often these \textbf{\textit{compensations}} were asked with the words of `support', `donation', and other similar words.
Thingiverse platform also supports `tip designer' feature that audiences can voluntarily send their credits, 
and creators who want compensation might enable this feature.
Our observation led us to design a detailed data analysis plan, to understand behavioral patterns of creators in compensating seeking, that we will detail more in the following section.
Second, to gain insights about people's experiences and honest thoughts of participants, we also collected existing conversations around `Thingiverse tip designer' from Thingiverse forum as well as Reddit,
and followed traditional qualitative research methods to extract patterns and insights.

\subsection{Analytics on Thingiverse Data: Data Collection and Inclusion Criteria} It has been observed that creators may seek compensation by either (1) posting external URLs that lead to certain financial transactions or (2) enabling `tip designer' feature in their designs, which is published as a web document in Thingiverse. 
Therefore, we started by retrieving relevant designs that are posted publicly on Thingiverse, among all possible data available by May 2020
(n=1,770,286, unique thing\footnote{We will refer each design entry as `thing' following the Thingiverse community convention, with its unique ID as thing number (\#).} ID from 2--4,360,000, excluding private designs and designs with 404 response due to post deletion), then classified into three categories below, also summarized in Table~\ref{tab:count}: 
\begin{table}[]
\small
\begin{tabular}{p{0.1\linewidth} p{0.14\linewidth} p{0.14\linewidth} p{0.14\linewidth} p{0.14\linewidth}}
\hline
& Donation (C1) & Webstore (C2) & Affiliate Links (C3) & Total \\ \hline
Thing Count   & 169,545   & 7,102  & 5,886 & 176,716  \\
  & (9.57\%)  & (0.40\%)& (0.33\%) & (9.98\%)\\ \hline
Creator Count & 13,387& 2,361  & 1,828 & 16,390   \\
  & (3.67\%)  & (0.64\%)   & (0.50\%)  & (4.49\%)  \\ \hline
\end{tabular}
\caption{Frequency and ratio of things and creators in each category (Total thing count: 1,770,286; Total creator count: 364,755)}
\label{tab:count}
\end{table}

\vspace{1.0ex}
\textit{\underline{C1. Designs Seeking Donation or Tip.}}
A thing (i.e. a design on Thingiverse) is considered as donation seeking if the creator of that design (1) enabled 'tip Designer' function on Thingiverse, or (2) explicitly asks for donations in the thing description and provides some methods for users to donate, such as links to the creator's PayPal, buymeacoffee account, or bitcoin address.
First, we retrieved designs with `tip designer' turned on using Thingiverse API (i.e. return creator.accepts\_tips field is set to True in API response, with the given unique thing ID), and it returned 163,069 things (made by 11,545 unique creators). 
To discover designs asking donation in the thing description, we collected all designs that contain keywords `donate', `tip', `support', `gratuity', and retrieved domains of URLs that are in the description, 
and merged hostnames of same websites to get unique websites. 
We finalized seven major donation sites, including paypal.com/paypal.me, patreon.com, buymeacoffee.com, yandex.ru, ko-fi.com, tipeee.com, and gofundme.com. 
Due to the variety of URLs, we only considered the major donation or money transfer sites, and the hostnames that appeared less than 50 times are not included. This process discovered 12,561 designs.

To cover more potential donation seeking designs that do not have URLs in the description or do not have 'tip designer' function enabled,
we performed text classification on thing description using fastText\cite{fastText} to find more potential designs. 
We collected the textual description in things containing the 7 domains mentioned above. 
Creators often display their intention of seeking donation by the sentence like "If you like this, please donate/buy me a coffee via [e.g., paypal/patreon URLs]". To train a classifier, we extract 1,552 sentences right before the URLs of two major sites PayPal and Patreon to be used as training data. 
By adding the same amount of random sentences from other random designs on Thingiverse as the negative cases, the text classifier is trained and used to detect additional potential donation seeking designs, which returned 35,469 more potential these things.
To validate these potential results, we used an interactive learning method similar to an existing work for image processing \cite{interactive_label} to bring human intelligence in the loop and iteratively update the results\footnote{For each loop, we reviewed and labeled only the top 20 of the results
, and these 20 labeled results will then be used to update the model. The model will then run again on the detected results and remove negative cases. This process repeats until all results are either verified by human or removed by updated classifier.}, which returned 3,437 new donation seeking things found. Summing all these designs we discovered and removing duplicates, we finalized 169,545 donation seeking designs in total.

\vspace{1.0ex}
\textit{\underline{C2. Designs Promoting Merchandise Sales or Services}}
A thing is considered as compensation solicitation if the creator (1) sells his/her design, product, or related works online on e-commerce web stores (e.g., Etsy) or personal websites, or (2) gets compensation by directing people to print design posted on 3rd party 3D printing services (e.g., Shapeways or Sculpteo) per each successful order. 
We discern this case based on whether the customers will be guaranteed to receive additional rewards, including finished prints as product, and other services.
Notice that some subscription services, such as Patreon, also provide exclusive content for patrons. However, they are included in C1 but not C2 due to the fact that people can still donate a random amount of money, and those who donate an amount less than the subscription level may not be guaranteed to get additional content. \par
For this case we used a similar approach, where we started by collecting sentences before Etsy and Shapeways URLs in description as these are two common web-store and 3rd party printing service, then trained a text classifier to discover more potential designs in this category. 
The classifier detected 31,820 designs. Using the similar validation approach using human-in-the-loop, 409 out of 31,820 are confirmed to be in this category.
We then retrieved domains of URLs, manually investigated their identity and service types, then classified into (i) webstores (Etsy, Cults3d, Myminifactory, Kitronik, RatRig, Wargaming3d, Myshopify, Miiduu, TheShop, Makershed, Gumroad, Fiverr, Cgtrader, 3d-mon) and (ii) 3rd party 3D printing services with original designs (Shapeways and Sculpteo).
We finally extracted 7,102 things in this category in total.

\vspace{1.0ex}
\textit{\underline{C3. Earning from Affiliate Links}} \label{sec:aff}
It is also observed that creators may advertise affiliate links via shared designs on Thingiverse, as they 
can get certain percentage of sales as payout for each successful order placed through these links. 
We put this in the separate category because creators do not sell or give additional products back to customers.
Shopping through the affiliate links is more about a 'kind' behavior instead of exchanging money to product or direct donation for contents.
Due to the diversity of format, we picked four most popular online shopping sites found in things ( based on URL count) that support affiliate links - Amazon (count: 27719, top 3 among all hostnames of URLs in description), Aliexpress, (count: 10801, top5), Banggood (count: 6089, top 6), and eBay (5915, top 8).
We first found a pattern in URLs which indicates affiliate links
by detecting a specific format, for example, `amzn.to' link and `amazon.com' links with `tag=' indicating the link is affiliate links. Similarly, `rover.ebay' indicates eBay compensates certain accounts with successful purchase order. For Aliexpress, we collected things with 's.click.aliexpress' or links that contain `aff\_platform=', and for Banggood, we collected URL that contains `p=' in it. 
This discovered 5,886 unique things in total\footnote{Count for each website (i.e., has duplicates for things contain URLs for more than 1 site): Amazon: 4,185, Aliexpress: 1,072, Banggood: 755, and eBay: 490 things respectively}. 


\subsection{Online Survey: Participants, Procedure, Responses}
Aiming to get a deeper understanding about honest thoughts, motivation, and expectations, we created an online survey asking a variety of questions covering their motivation of sharing models in the open market, the methods for getting the compensation they have tried, the satisfaction of the current profit and the expectation, and how compensation from all sources help creator's continued creative labor and life. 
All questions were left optional in case participants may not want to reveal certain information.
We randomly selected 700 creators from all the creators of things we collected in the above three categories and sent them the online survey invitation via private messages. To avoid biases, we intentionally did not choose from active creators (i.e., we ignored known information in the recent log-in record). Participation was completely voluntary, no compensation has been promised. 44 forms were returned (6.3\%), 
with 43 valid responses left by removing a duplicate submission. 


\section{Results \& Finding}

    
    

\subsection{Frequency, Rate, and Trend: Compensation Seeking Designs over All Designs}
\begin{figure*}[t]
    \centering
    \includegraphics[width=\textwidth]{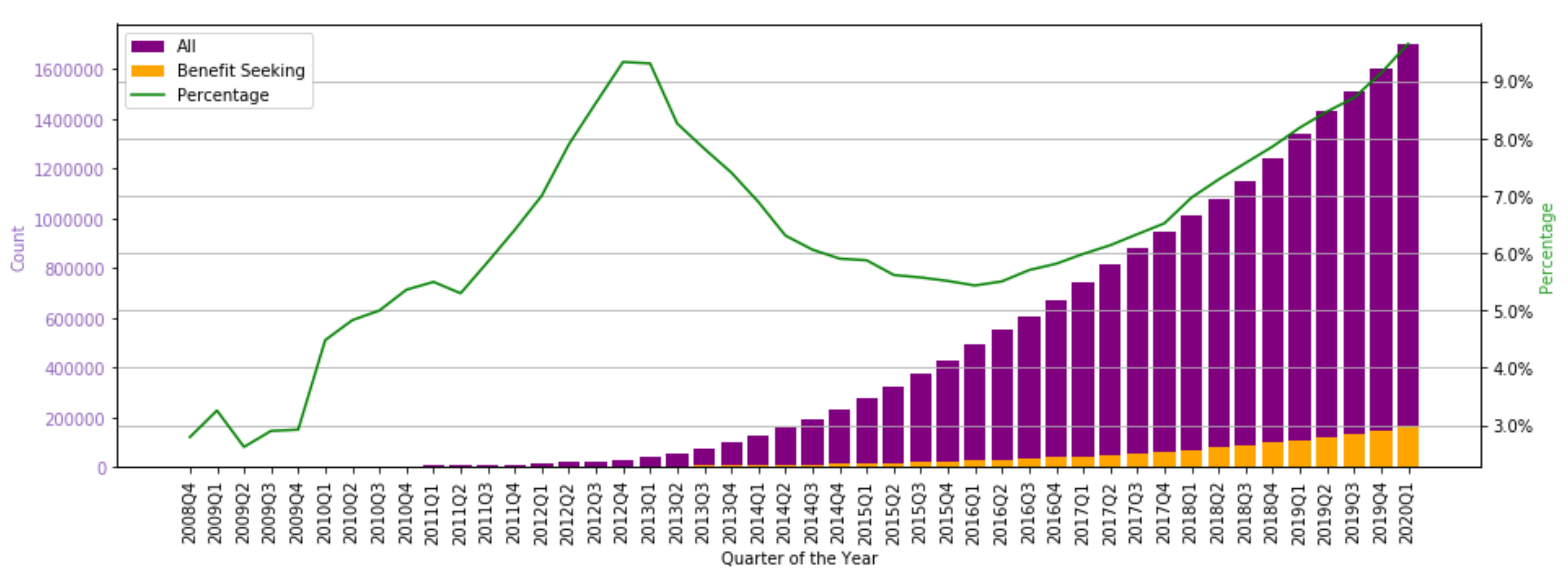}
    \vspace{-2.0ex}
    \caption{Cumulative count of all things and compensation-seeking things and percentage of them by a quarter of the year. As the data for Q2 2020 is not complete, it is not included in the graph.}
    \vspace{-2.0ex}
    \label{fig:count}
\end{figure*}

In general, at least 9.98\% (176,716 out of 1,770,286) designs showed some intention of seeking compensation for their creative labor.
Among creators who seek donation (C1), 86.2\% of them use the built-in `Tip Designer' function, and 21.03\% of them asked for donation or monetary support, for example, by adding a PayPal account to the description field. 
7.27\% (n=974) of them utilized both methods to ask donations. 
For all creators who seek compensation, 
7.23\% (n=1,186) of them utilized at least two of the three methods (C1-3).
We also retrieved the dates these things were posted to Thingiverse, and counted the number of them compared to all designs (October 2008 - March 2020, per quarter), to see the trend. 
As shown in figure \ref{fig:count}, the percentage of compensation seeking things has been increasing, with the only exception near 2012 Q4 where the proportion diminished. 
This could possibly due to the rapid increase of non-compensation-seeking designs, especially customized designs from existing designs, as we also observed that the first customized thing made by Makerbot Customizer was posted on January 08, 2013, which is very close to the starting point of the decreasing trend of percentage of compensation seeking designs on Thingiverse.
Except for this case, we can see the compensation seeking behavior is also getting very popular.  

The count of affiliate links also shows the popularity of utilizing this method to get additional compensation. 
There are 5,886 unique things in total that contain affiliate link, where 4,185, 1,072, 755, 490 things have Amazon, Aliexpress, Banggood, and eBay affiliate links,respectively. 19631 things contain Amazon links in description, and 21.31\% (4185 out of 19631) are things that contain Amazon affiliate links. This percentage is 18.03\% (755 out of 4186), 16.15\% (1072 out of 6635), 8.35\% (490 out of 5862) for Banggood, Aliexpress, and eBay, respectively. This shows that using affiliate links is also a popular choice for creators to receive extra compensation, both explicitly (i.e. mention in description that links are affiliate links) or implicitly (i.e. directly add them to description without saying they are affiliate links).

\subsection{New Opportunities Featured by 3D Printing Community: How Creators Make Money?}
Besides accepting tips using built-in Thingiverse function, there are various methods of getting compensation that creators can use to receive additional earnings, which could be advertised by their shared 3D contents via open platforms.

\subsubsection{Type 1. Subscriptions and Paid Memberships for Exclusive Contents}
One common way that creative workers gain additional earnings from their content and labor is through a monthly subscription or a paid membership, and it also appeared to be a very popular method among creators on Thingiverse.
One typical site of this subscription-based donating mechanism is Patreon. 
It is one well-known platform to support indie-creators \cite{5PlftformsCreators:online}, and also served as a platform for supporting designers and providing subscribers (also called 'patrons') with exclusive contents and rights; and it claimed to have paid out over 2 billion dollars to Patreon creators since 2013 \cite{RN5}. 
Although audiences have the freedom in setting amount and method for supporting each creator, they often pay the membership fee at pre-determined levels set by the creators. 
Fees also vary, from \$1 per month for entry level and \$10 per month for a more advanced level, to some even higher levels that are set by creators. 
Subscribers can get rewards, including exclusive content, access to the community, and private services such as consulting, depending on the level of subscription they chose. 
Many creators on Thingiverse put the link to their Patreon creator profile in the description section (7,799 things in total), to encourage people to visit and discover more similar contents and high quality contents (e.g. high resolution model). \par
\begin{figure}
    \centering
    \includegraphics[width=\linewidth]{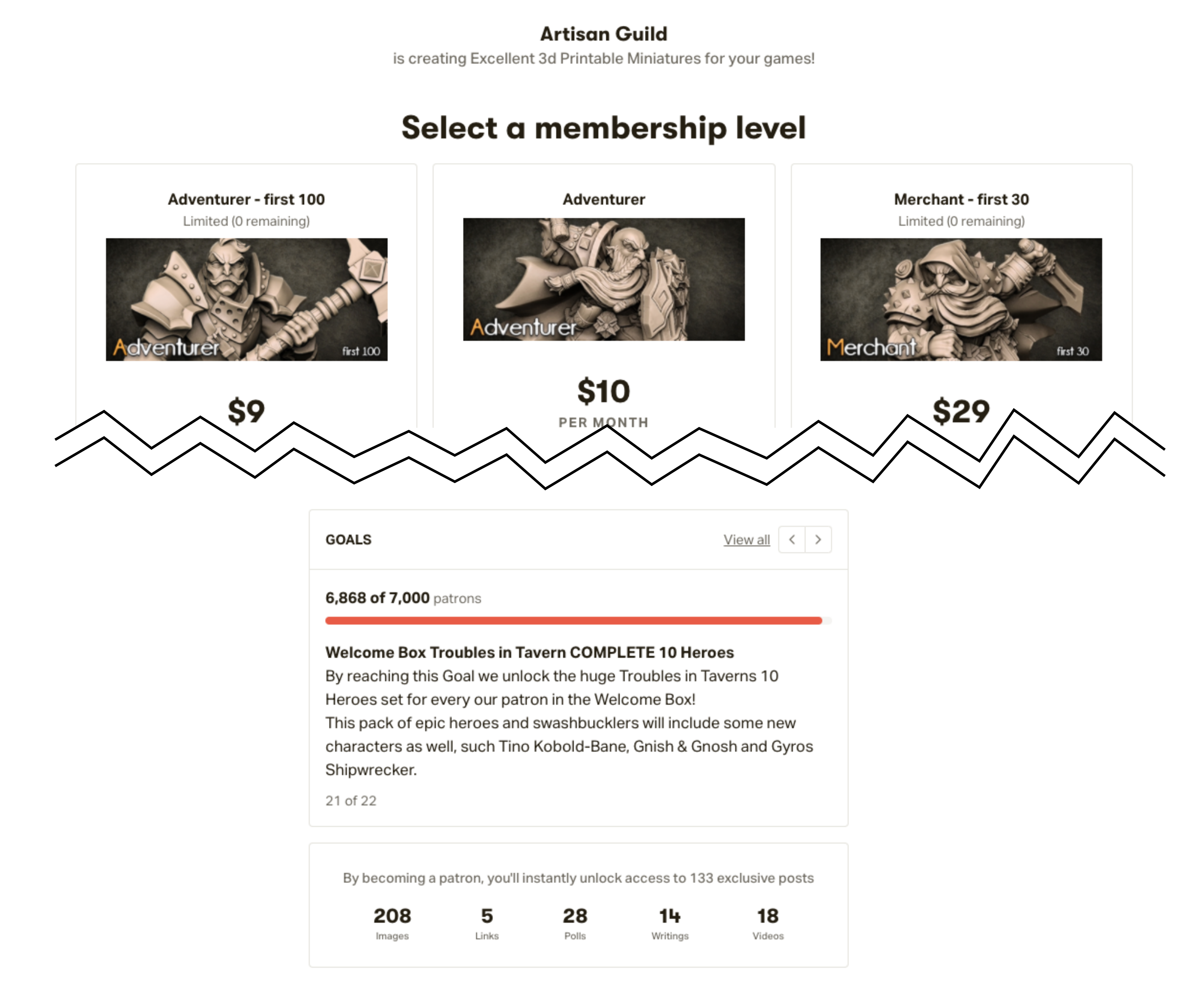}
    \caption{Example creator's Patreon profile page. Many creators set membership levels, set various goals, for subscribers to know more about their increasing benefits and support creative endeavors}
    \label{fig:patreon_profile}
\end{figure}

Although Patreon does not represent all subscription business, as it is one the most popular creator-supporting service found in Thingiverse (by count), we chose this to understand how Thingiverse creators utilize subscription method to compensate their work.
We retrieved all Patreon links in the description of all designs on Thingiverse, and identified 414 unique Patreon accounts.
Excluding invalid accounts (e.g. returns 404  or account under review)
, we found 264 unique Patreon pages.
We then manually collected details about their Patreon page, including membership levels, membership benefits, goals, number of patrons, and current monthly donation, as an example shown in Figure \ref{fig:patreon_profile}.
These fields are all optional depending on user's setting. Based on the available data, we analyzed the distribution of membership levels, and the type of extra benefits that subscribers can get from the creators.\par

\textbf{Creator Profit} Creators on Patreon, though not required, can have predetermined membership levels and the corresponding benefits on their page. 
Subscribers can choose from one of these membership levels or donate a random amount of money.  
Of 264 unique creators, 239 of them have pre-determined membership levels ($mean=\$21.35, median=\$7$). 
Some creators may set a very high subscription level (e.g., \$1,000 for the highest subscription level), but majority of the subscription levels are under \$60, and thus the distribution of cost is right skewed.


\begin{table}[h]
\small
\begin{tabular}{p{0.15\linewidth} p{0.15\linewidth} p{0.5\linewidth} }
\hline
\textbf{Rewards Types} & \textbf{Creator Count} & \textbf{Examples} \\ \hline

\rowcolor{lightgray}Exclusive content & 177 (67.04\%)       & Member-only designs, weekly updates\\ 
Community & 162 (61.36\%)               & Joining members community (e.g. Discord), polls and votes for next design, early access, FAQ, chat with creators\\ 
\rowcolor{lightgray}Licensing & 71 (26.89\%)                & Permission to sell prints (but not original design) \\ 
Credits / appreciation & 47 (17.80\%) & Name on YouTube video, gifts, personalized Thank You video/audio.\\ 
\rowcolor{lightgray}Discount & 44 (16.67\%)                 & Discount or free access in creators' own web stores (e.g. Etsy, Shapeways) \\ 
Additional format & 35 (13.25\%)        & Editable source file (original design and source code), high resolution files, modified version of free designs \\ 
\rowcolor{lightgray}Personal service & 35 (13.25\%)  & Consultation, mentoring, accept personal modeling request, modeling/design lessons and tutorials\\ \hline
\end{tabular}
\label{tab:patreon_benefit}
\caption{Classification of Patreon Membership Benefits. \% refers to ratio of 224 creator profile.}
\end{table}
\textbf{Member Rewards} In return to patrons' subscriptions and donations, creators may provide additional rewards to these subscribers based on levels. This covers various benefits from exclusive content to one-to-one consultation. 
Also, creators may release new content monthly instead of all at once, to get a high retention rate of the patrons to make the income from subscriptions stable.
In general, the higher the level, the more benefits the patron can get, such as limited edition or personalized design. 
To understand customer's motivation to pay extra money to creators with higher membership, we summarized these benefits that are found from creators' profile pages as shown in the Table 2.
Exclusive content and community related benefits are found out to be the two most common benefits, which shows that exclusive content and more involvement with creators are relatively more attractive to subscribers who are willing to pay creators.
The result also shows that creators are willing to sell \textit{services} (e.g. provide more personalized benefits) instead of providing \textit{products} (e.g. exclusive designs) as this might result in one-time payment, and utilize their skills and expertise to attract more patrons and keep stable revenue.


\subsubsection{Type 2. Merchandise Sales Related to 3D Design}
Another common way of gaining extra earning through 3D printing is to sell finished prints of models (for those who have no easy access to printers), prints in special materials, assembled products with 3D printed parts and other components, or a kit of parts/components that are used in the design.
Things that fall into this category usually have links to the creator's web stores (e.g., Etsy) in their designs' description part on Thingiverse, or links to the creator's printing store on a 3rd party platforms (e.g., Shapeways) to receive commission for their design from successful print orders. 
We categorized three main patterns as follows:

\begin{figure}[t]
    \centering
    \includegraphics[width=\linewidth]{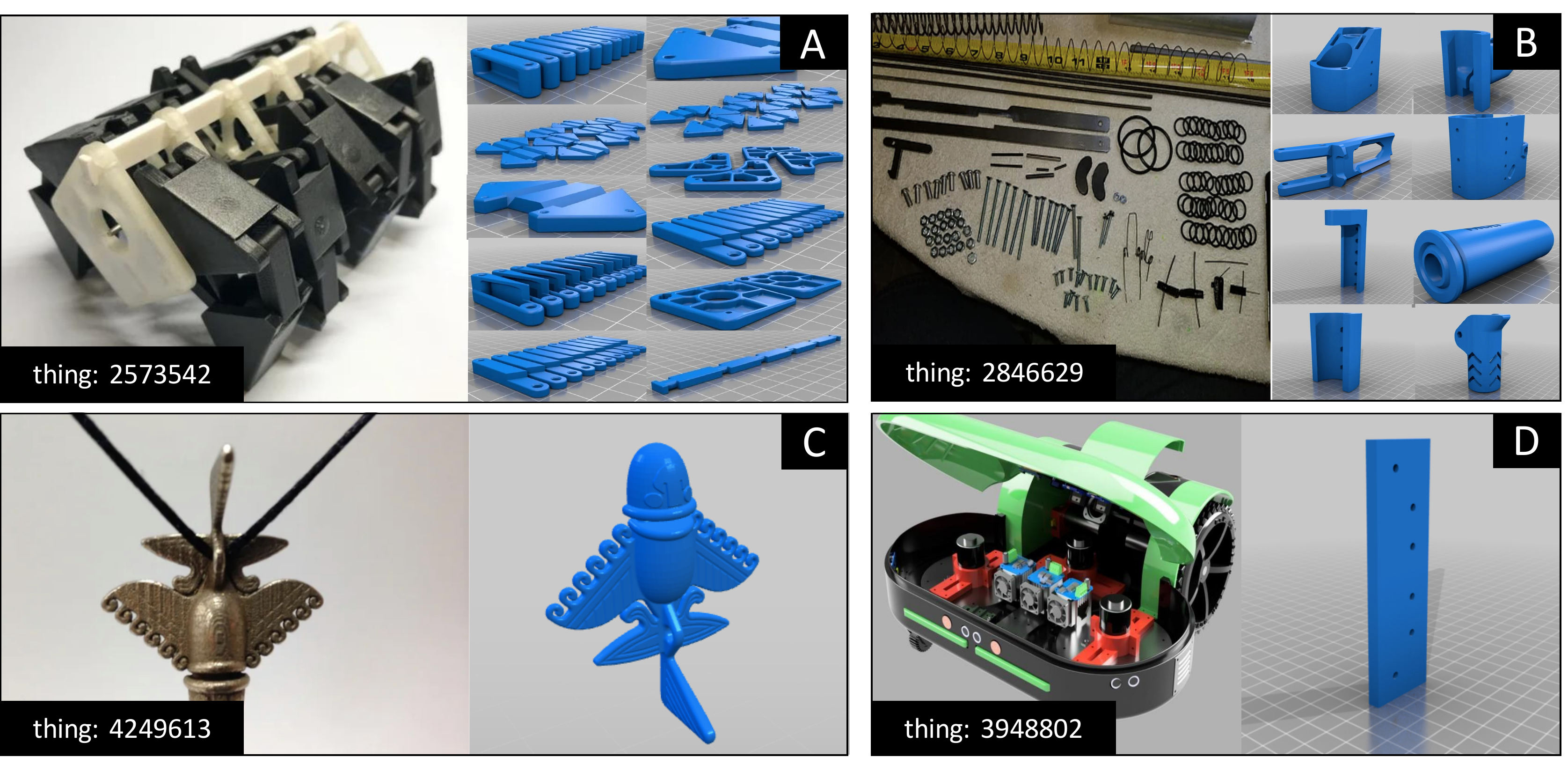}
    \vspace{-2.0ex}
    \caption{
    Example in each classification of compensation type (A) full assembled design and all required components, (B) selling hardware kit that are required to assemble components in 3D,
    (C) full print via 3rd party printing service using special materials, and
    (D) full set of models and components where opensource design is only the small part.}
    \label{fig:pay_for_full}
\end{figure}
\textbf{Finished prints} 
This is the most frequent case, where creators include the URL to their personal web stores (e.g. Etsy, Gumroad, etc.) in their designs' description on Thingiverse. Creators get financial gains from their printing capability in addition to the designs by selling finished prints of their own designs.

\textbf{Final Products with Print and Assembly} Many creators share complicated designs that require intensive assembly, and some creators may sell assembled products in store so that people can directly purchase a ready-to-use product. 
For example, a mechanical walking toy (Thing\#2573542, Figure~\ref{fig:pay_for_full}A) consists of 11 files. 
Audiences have the choice of either printing them all and assemble them, or directly purchase an assembled one from creator's store on Etsy. 

\textbf{Kits and Tools} While for some other designs, creators usually sell a kit of the design instead of the assembled products, and this can be found in many DIY designs, especially for those that require additional components that cannot be easily printed, such as electronic components or metal parts. One example is a dart blaster (Thing\#2846629, Figure~\ref{fig:pay_for_full}B). 
The design requires both the 3D printed part, such as shell and handle, and other additional hardware components, such as spring and screws. 
Users can either print the necessary part and buy other hardware components separately, or buy a kit that contain all hardware needed to complete assembly, which could potentially save time and expense on purchasing the right components for the design and also support the creator's work. 


\textbf{Finished Prints through 3rd-party Printing Services}
It has also been found that creators may direct users to print their designs through 3rd party printing services, such as Shapeways and Sculpteo. 
Creators have a virtual `printing shop' hosted by them, and can receive a commission for each successful order through their personal shops (Figure~\ref{fig:pay_for_full}C). 
Customers often utilize these special services due to availability for special materials using high-end printers that are not easily printable using low-cost printers, including soft plastic (Thing\#3733220), metal (Thing\#2736009), and ceramic (Thing\#150087). 
This gives creators more possibilities and choices in their designs, as they can create designs that usually require special materials, such as a jewelry ring (Thing\#1230141).\par

\textbf{Full Version of Incomplete Design}
Some creators post limited or incomplete version of designs on Thingiverse for public, then sell full design in their personal stores (Figure~\ref{fig:pay_for_full}D). 
One example (Thing\#4249613) is a set of miniatures, where one of them are posted publicly on Thingiverse, and the full set (which contains 3 models) can be purchased at cults3d. 
Another example is a single design (Thing\#3948802). One part of the design is on Thingiverse, but the full design (which has more model files) needs to be purchased in store. Images of these examples are shown in figure \ref{fig:pay_for_full}.
Other examples include a torso of naked body, where the full body sculpture needs to be purchased (Thing\#2514495).  
Users can decide whether they want to pay to get more variation of the design or not.

\subsubsection{Type 3. Consulting/Personal Design Task}
Many creators also provide paid personal services in the field of 3D printing, such as 3D design on-premise through Patreon, or other freelancer platforms (e.g., UpWork, Fiverr). 
As known from above, 13.25\% of Patreon pages on Thingiverse proposed personal service. 
Of these pages, 74.28\% of them offer on-demand modeling service, providing the modeling or crafting, including
customizing existing models based on customer's special requirements, and 54.28\% of them provides consultation or tutorials, including technical support, modeling tutorials, and consultation on personal projects of customers. 
This shows an additional clue of compensation can be received not only from the physical products or models, but also from creator's \textit{professional skills}.\par
\begin{figure*}[h]
    \centering
    \includegraphics[width=\linewidth]{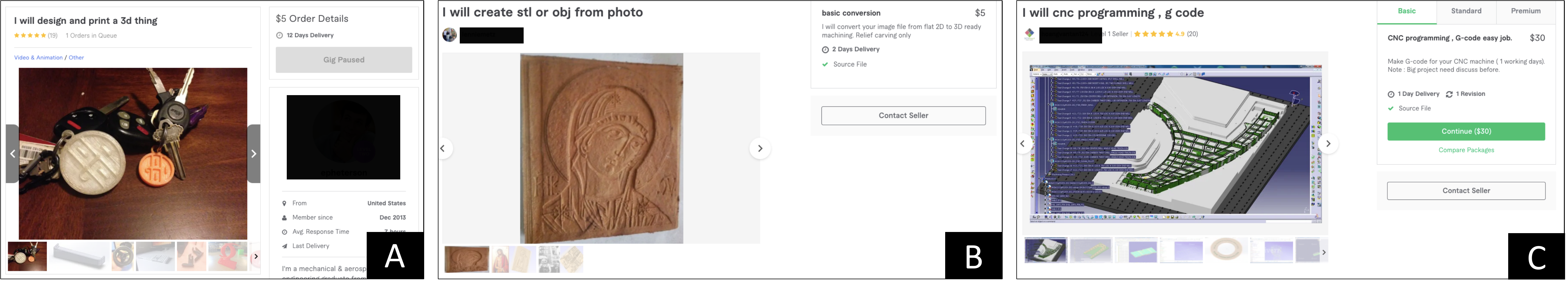}
    \caption{Example of creators trying to get paid by their professional skills, advertising through shared 3D contents: (1) 3d printing, (2) converting photos, and (3) gcode programming, etc. Creator bio and information is removed.}
    \label{fig:fiverr}
\end{figure*}
Another platform Fiverr, which is also popular among creators who seek compensation, focuses more specifically on freelance services. 
51 things explicitly mentioned Fiverr in the description, and 20 unique Fiverr links (excluding links that belongs to the same creator, or links that are no longer accessible) were associated with designs on Thingiverse. 
Similar to the personal services on Patreon, creators also provide various paid personal modeling, programming, printing, or drafting request, such as non-specified general modeling service (Figure~\ref{fig:fiverr}A)
, specified modeling service (e.g. create STL model from photo, Figure~\ref{fig:fiverr}B)
, cnc programming and g code (Figure~\ref{fig:fiverr}C)
, and technical drawing service for manufacturing, and more.

These instances with paid personal services demonstrate that there are creators who are compensated for their expertise and individual tasks in addition to contents that are shared in open platforms-- similar to utilization of multiple digital platforms in existing online gig economy \cite{GigEconInfo}.


\subsection{Type of Contents: What Things Creators Make Money from?}
While there are more people who believe in the promising future of the 3D printing, there are also people with curiosity why people would pay for 3D designs, instead of buying mass-produced products from Amazon. 
As we see in many applications that the variety can be better accomplished via customization, in that, which 3D printing has shown its promises, it is important to know what type of products has been gaining its popularity particularly to make profits.
To understand the type of things, we retrieved the information about thing `category' and thing `tags' of designs.

\subsubsection{By Category: 3D printing, Hobby, Household Items, and Toys \& Games}
To publish a thing on Thingiverse, creators are required to select the category from a predefined list, and the platform provides two levels of categories to help creators best describe their work, such as Fashion, Gadgets, Household as the top level category, and Costume, Camera, Kitchen\&Dining as subcategories. 
We retrieved the first level category of set by Thingiverse, which include 11 different categories: `3D Printing' (19.17\%), `Household' (18.62\%), `Hobby' (12.51\%), `Art' (10.76\%), `Fashion' (9.68\%), `Toys \& Games' (7.18\%), `Tools' (5.96\%), `Gadgets' (5.6\%), `Other' (4.65\%), 'Models' (4.4\%), `Learning' (1.47\%).

We retrieved the top 7 categories with more than 10K instances, as shown in the Table~\ref{tab:popular_category}.

\begin{table}[h]
\small
\begin{tabular}{p{0.08\linewidth} p{0.2\linewidth} p{0.1\linewidth}|p{0.08\linewidth} p{0.2\linewidth} p{0.1\linewidth} }
\hline
\multicolumn{3}{c|}{\textbf{Sort by count}}   
& \multicolumn{3}{c}{\textbf{Sort by ratio in each category}}   \\ 
   \textbf{Order} & \textbf{Category} & \textbf{Count} & \textbf{Order} & \textbf{Category} & \textbf{Percent*} \\ \hline

    1  & 3D Printing   & 32,823  & 1  & Toys \& Games & 19.53\% \\  
    2  & Hobby & 26,561 & 2  & Models& 19.01\% \\ 
    3  & Household & 26,008 & 3  & Hobby & 12.02\% \\ 
    4  & Toys \& Games & 24,758 & 4  & Gadgets   & 10.35\% \\ 
    5  & Art   & 16,600 & 5  & 3D Printing   & 9.53\% \\
    6  & Models& 14,779 & 6  & Learning & 8.86\%\\
    7  & Gadgets&10,248 & 7  & Art      & 8.47\%\\
    
    \hline
\end{tabular}
\caption{Popular thing category (with more than 10K designs)in each compensation type, sorted by percentage (*Percentage indicates the ratio of things with compensation solicitation among all things in the same category)}
\label{tab:popular_category}
\end{table}

Considering the difference of total count of each category\footnote{ 
`3D Printing'
is meant to parts and accessories for printers, such as extruder, parts, and tests as sub categories, but somehow it is misunderstood by creators so anything can be 3D printed are often categorized into this. 
Thus, it is possible that the simple count does not necessarily represent its category, and we report the ratio of compensation seeking design in each category as well.}, we also retrieved the top 7 categories sorted by relative percentage in that specific category (i.e. percentage of compensation seeking designs in each category). As shown in Table \ref{tab:popular_category}, `Toys \& Games' become the most popular, where people may post and sell models and miniatures of games and movie characters, such as the collection of DnD (Dungeons \& Dragons) miniatures (thing\#3054701).

\subsubsection{By Thing Tags: ender\_3, fpv, customized, dnd, stencil, miniature, and camera}
While optional, creators can also add tags to their design to help audiences quickly search, and combinations of user-defined tags often better accurately describe the identity of the design.
Thus, we retrieved tags of compensation-seeking things in the top 7 categories sorted by count in Table~\ref{tab:popular_category}, and extracted the top 10 tags from each category. 
\begin{table}[h]
\small
\begin{tabular}{p{0.2\linewidth} p{0.7\linewidth} }
\hline
 \textbf{Category} & \textbf{Top 10 Tags}  \\ \hline
\rowcolor{lightgray}3D Printing & ender\_3, creality, filament, spool\_holder, prusa\_i3, filament\_spool\_holder, extruder, anet\_a8, prusa, 3d\_printer
    \\ 
Hobby & fpv, drone, quadcopter, mount, arduino, case, diy, holder, customized, airsoft
    \\ 
\rowcolor{lightgray}Household & customized, holder, vase, decoration, kitchen, container, christmas, box, cookie\_cutter, household
    \\ 
Toys \& Games & dnd, tabletop, miniature, dungeons\_and\_dragons, wargaming, rpg, warhammer, boardgame, 28mm, pathfinder
\\
\rowcolor{lightgray}Art  & stencil, customized, art, sculpture, logo, sign, 2d\_art, openscad, statue, decoration
\\
Models & miniature, tabletop, model, warhammer, animal, warhammer\_40k, wargaming, 28mm, prop, toy
\\
\rowcolor{lightgray} Gadgets & camera, iphone, stand, case, holder, mount, gopro, phone, phone\_stand, smartphone
    \\ \hline
\end{tabular}
\caption{Top 10 thing tags in the top 7 compensation-seeking categories (with more than 10K designs, sorted by total count of designs) 
}
\label{tab:popular_tags}
\end{table}
As shown in table \ref{tab:popular_tags}, popular tags for 3D printing are related to 3D printer parts, printing equipment, or materials. 
The popular tags in both `Toys \& Games' and `Models' show that a number of compensation-seeking designs are highly related to video or tabletop game/movie/anime characters and miniature.
This might indicate a potential popularity of fandom culture in compensation-seeking designs among creators. 
For art category, tags are related to both 2D and 3D designs, which also shows potential opportunities of customized on-demand design request.
For Hobby and Gadgets categories, the popular tags are often related to DIY projects, parts, and accessories, that could also potentially lead to selling kits, parts, and assembled products from web stores, as we found above. 
These categories are also popular in things with affiliate links, as things may require additional components or devices (e.g., Arduino, electronics, and small components or materials), which are usually available for purchase online; and thus the creators can put affiliate links in description both for helping other makers to quickly locate the parts needed and get a compensation from successful orders-- which might be motivation of designers in creating 3D models; create and share 3D designs that audiences download and buy required things via provided channels.


At the same time, it implicitly tells us that certain categories such as `Fashion' (jewelry, accessories, etc.), `Learning' (biology, math, etc.), `Tools' (chess, construction toys, etc.) are either not very popular among creators who are interested in sharing contents for profit, or possibly, creators who design and share those things do not seek compensation through designs shared for public. However, these two hypotheses deserve another in-depth analysis.
\par


\subsection{Motivations, Expectations, and Gaps}

We further conducted a survey with Thingiverse designers to ask their motivation in sharing designs, their expectation of compensations, and the current situation. We sent out 700 survey invitations on Thingiverse through private messages, and received 43 valid survey responses. All responses are anonymous and optional.

\subsubsection{``Share and Share Alike, Everyone Wins.''}
We started the survey by asking creators their motivation of sharing designs for free. We provided 4 pre-set options for this question, including (1) I want to share the talent I have with the community, (2) I take this as accessible storage of my work, (3) I want to share other 3D designs to get more attention through this work, and (4) I want to advertise my other creative work portfolio (PodCast, Illustrations, YouTube, etc.). 
In addition, participant can also enter a custom response in case none of the above options applies. 
Of 11 responses (excluding who skipped to answer), 8 responses (72.7\%) are for altruistic reasons, including (1) and two customized responses \textit{``share and share alike, everyone wins''} and \textit{``I want to make the world better with my designs''}. 
Other options (2)-(4) got one response each. The result shows that although we are targeting the user group that indicated at least one method for receiving compensation, the majority of creators who responded still share their designs for free, primarily to contribute to opensource community. \par

Creators may post their designs for fun or mostly for altruistic reasons, such as \textit{``giving back to 3D printing community''} (R2), \textit{``helping others''} (R4), and \textit{``keeping the community going''} (R36). 
Some creators turned on the 'tip designer' function because other users requested it to show their appreciation, or because creators may want to refund from initial investment. For example, one mentioned that \textit{"I gave the possibility to leave a tip on Thingiverse for my designs. I was hoping at least to refund some material such filaments or 3D printer parts. I did not want to gain anything from sharing the designs. I believe that the 3D printing community is great because of the free availability and sharing of Knowledge and Ideas."}(R8)
One participant also mentioned that \textit{"If I don't need the money for the time I spend on that creation, then I won't require it. I want everyone to have the ability to create regardless of their ability to pay for it, but I understand that other creators need to be compensated for their time so they can continue creating."} (R25)\par
\subsubsection{``I Would Like to Make Someone Cares.''}
While altruism could be yet the primary motivation, it can be found that compensation could also bring additional motivation for creators to post more designs. 
For example, one participant mentioned,
\textit{``Sometimes even a very tiny small tip would encourage me to make/publish 3D models.''} (R43) 
Giving tip is also a way of showing appreciation, as one quoted \textit{``Because it's difficult time, just 5 bucks would be awesome to show that people care what you do.''} (R18)
Insufficient compensation may let creators feel that the time they spent on designs are not well acknowledged. The participant added \textit{``If you make something and people do not care, it just takes down the mood to upload new thing, like the motivation will just disappear.''} (R18) Thus, compensation would be another motivation that keeps creators sharing more designs, including those are free for the most of audiences on open market.

\subsubsection{``I Want Compensation from what I've Spent Time On.''}
Creators' expectations of compensation are diverse, depending on the type of models or the actual need of compensation. 
We asked the compensation amount that creators expect to receive from their designs that are shared online, and the range of responses is quite vary, from as low as \$10 in total to \$1,000 per month. 
This may also depend on how much time they spend on creating designs. 
For example, one participant explicitly mentioned that he/she wants to \textit{``get a job and income from my designs.''} (R12)
Other creators might take this as freelancing job, doing is as a side hustle which can bring \textbf{extra} income as also seen in many existing online gig economy such as Uber and TaskRabbit \cite{GigEconInfo}, as participants quoted \textit{``This is fun, but I can stop whenever I want and move on to other things. Once it’s my main income I’m trapped. Keeping it a low key side hustle takes a lot of pressure off and let’s me create better because the motivation is love of the craft and not making rent.''}  (R26)
Some also mentioned that they only expect compensation for \textit{``the ones I've spent the most time on''} (R30) or complicated designs only, while keeping some other designs freely accessible.
\textit{``If I don't need the money for the time I spend on that creation, then I won't require it.''} (R25),

\subsubsection{``I Had High Hopes but Everyone Wants Us to Work for Free.''}
Although creators have various expectations, there is one common mind shared by creators, which illustrated the gap between the expectation and reality, as one participant mentioned \textit{``Thingiverse is now only used to show designs, I have never received tips from using this platform. I had high hopes but everyone wants us to work for free.''}  (R41)
On average, the total amount of compensation that participants indicated that they have received is \$9.32 per creator, and 20 of 34 (58.8\%) have not received any.
We also asked participants whether the current compensation they received could motivate them to post creative works full time by choosing between 1 (Strongly Disagree) to 5 (Strongly Agree), and on average the score is 2.23, which shows that the current amount they received is still below their average expectation. 
In addition, none of the participants think the current compensation is a stable source of income. 
This gap may potentially discouraged creators to post more or quality designs, as one participant mentioned \textit{``I've stopped making models altogether, It's just not worth my time. People expect everything to be free these days.''} (R37)
It may also cause creators to move to paid sites to seek enough compensation, as another participant mentioned \textit{``I am strongly considering moving to a paid site where I can receive a few dollars per download. Thingiverse tips are less than \$100 for a complex design with over 15,000 downloads."} (R34)\par
Besides the amount of compensation in total, another reason which may stop creators from spending more time on creating and sharing design is the instability of the compensation. 
28 out of 42 participants agree or strongly agreed that the compensation they received is too irregular to be a steady income that they can rely on. This may due to the nature of certain compensation method, such as one-time tip through PayPal or a one-time payment to unlock all designs (which could potentially be re-distributed online). 
From another point of view, this shows potentiality of subscription-based monthly donation, as the monthly compensation is easier to predict and thus the creators can plan the time they spend on creating new things accordingly. 
This could also motivate creators to post new designs in a planned manner, in order to keep high retention rate.\par

The results echo our observations in the existing discussion amongst Thingiverse users through forums, as they do not think voluntary tipping is enough to sustain it as day job: \textit{``I sometimes receive tips, but a regular \$10 a month would be utopian.''} (Thingiverse user DrLex), \textit{``If I'm wrong, and anyone here makes more than 10\$ a month in tips, I'd be surprised''} (Thingiverse user Drewrt).






\section{Discussion: Opportunity, Concerns, and Implications}

\subsection{New Socio-economic Opportunities Emerged by 3D Printing and Online Platform}
As shown in this work, 3D printing has been presenting many new societal impacts with the rapid growth of opensource platforms online.
With the open platform, the 3D printing community rapidly responded to the critical needs in personal protective equipment for clinical professionals during the recent COVID-19 outbreak \cite{NIHCovid}. 
Various group of people, such as America Makes and Make4COVID, participated in various pipeline of supply chain, as designer, design verifier, local manufacturers, quality assurance, and many more \cite{make4covid}.
On Thingiverse, `covid-19' tag also quickly became very popular, with its 3,556 frequency ranked at the 90th most popular tags among all designs (as of May 2020).
As we have seen through this work, 3D printing open market has also been used to advertise creative works and get attention, which enabled creators to 
use this opportunity to launch a new open-source based entrepreneurship via successful kickstarter funding.
Although we need further studies to identify creators' identities\footnote{We retrieved user accounts and looked into their public bio page to extract their self-reported identity.
However, due to limited selection category originally set (e.g., educator, students, maker) on the bio page, these are limited to know their occupation or current social position and more}, we also observed some empirical evidences about new societal opportunities for various people, including entrepreneur, retiree, female, and more.
Furthermore, known from the goals set by 139 creators who owned Patreon accounts (139 of 264 Patreon creators set goals on their Patreon page),
some creators also want to quit or decrease day job hours (5 out of 139), open a dedicated front store to sell more products at easy access (6 out of 139),
or even hire people (7 out of 139),
which we can expect even expanded job opportunities that 3D printing community can bring to our society, more to the underrepresented groups who needed more job flexibility or limited mobility, and more.
3D printing has brought more and more options for home workers, where we can see the future with expanded opportunity from DIY and craft to freelancing design and printing tasks.

\subsection{Tension between Opensource Culture and Copyrights to Credit Labor}
As introduced earlier, 
people have started sharing their own designs with altruistic motivation, and it helped the community grow and mature. The growth of communities with more people owning 3D printers at home also broadened the audiences who can participate in this new labor market.
Now the technology and culture is grown enough to craft new opportunities (jobs), for example, custom design in open repositories now can be completely 3D printed at nearly-commercial quality, intriguing customers buy these over mass-produced items.

However, potential concerns has emerged in various aspects, particularly with IP rights of shared contents. Protecting copyrights of creative contents online somewhat compensates the societal benefits, in particular, regarding reuse thereof for derivative creations\cite{kiskis2005compensating}.
Activism to make historic assets and intellectual property has been getting popular which could become beneficial for wider community, as example can be observed with 3D models of 3D scanned historic sculptures at Museums\footnote{Example available by Thingiverse user cosmowenman, with long story for law suit cases \cite{MuseumSculptures}}
The most controversial case caught its attention when people made profit by selling complete prints on Ebay, made of models that were originally shared for free by others \cite{eBaySeller}.
The designer shared for ``Non-commercial'' but a seller at the closed market such as Ebay downloaded a bunch of models and sold for price, with the wrong interpretation of Terms of Use, saying \textit{``Once you share for opensource, you lose the right''}. 
Since the most of those items were down and so unavailable from Ebay, some creators chose to take different actions, where some has begun to make earnings by selling `commercial license', officially allowing others to sell final prints if they pay fees to the original designer (e.g., Thingiverse user Robagon), while the other deleted his all free designs from open repository, leading audiences to visit his webstores hosted in Shapeways (Thingiverse user mz4250), where people can purchase printing service, but not the design file itself for free replication or remix.

\subsection{Potential Issues with Remix, Customize, and Sensitive Things}
Licensing may sometimes be unclear on certain websites for makers and designers to notice, particularly in regards to fork and renovating new features (remix) based on opensource libraries and designs.
It has been noticed that platform does not guarantee any \textit{equitable} compensation to the right holder, while, unfortunately, it may provide tools for abuse\cite{compensateCreativity}.
Currently, remix and customize are also one popular native function on Thinigverse, but what if remixes and customized design get compensation while the original is not? 
In reality, among all things that solicit compensation for some format, 7,012 are customized, where we might need more in-depth analysis if the majority of them were done by another Thingiverse users or the designer-selves. 
We have seen that people actually obtained tips from remixes even without complex re-design \textit{``I found which part I was tipped by, and it was something I remixed, just to flatten in out, because it was at a very difficult angle to print at. This had its beginnings with Ryan’s coupler, so thanks Ryan''} (Thingiverse user Jeffeb3), of which the original designer is unable to directly benefit from it as s/he did not enable tip designer or asked for donation at all.

\subsection{Towards Fair and Ethical Technology to Support Creative Endeavors}
There have been increasing discussions among creators, which platform they can trust to get paid.
\textit{``I never tip through thingiverse because makerbot clearly doesn't deserve 30\% cut they take.
Also secondly, I have evidence that they just steal some tips (some people told me they sent me a tip through thingiverse, the paypal accepted the transaction but I didn't get anything out of it).''} (Reddit user HairyBeardman)

Recently, Thingiverse has allied with 3D hubs\cite{3DHubs}, which helps customers without 3D printers or expertise/time to 3D print.
Users can place a print order directly to 3D hubs, out of free 3D models on Thingiverse. 
Since then, creators questioned \textit{``Who’s making money off my design?''} (3DHub user Alexander\_Gaarn) that ignited a long debate among stakeholders. 
While some people believe that \textit``{To me, the entire point of designing things is to prove that 3d printing can be useful and not just flashy. I just wanted to make sure that no one was trying to undermine the altruistic endeavor.''} (3Dhub user Alexander\_gaarn), others may arguing back \textit{``If you are mad at people here because someone took your work that you `gave away freely' and hired one of us to print it for them, your anger is severely misplaced. If someone bought a printer so they could print your design, would the manufacturer of the printer also be `making money off your design'?''} (3Dhub user MindFull). 
Nonetheless, the Thingiverse and 3Dhub officials claim that they \textit{``enable 3D Hubs by default \textbf{only} for designers who’s licenses allow for commercial use and we make it easy to disable''} (The full conversation available at \cite{Thingiverse3Dhubs}).
The controversy could be alleviated by transparent information, while we need to invest more efforts into developing ethical and fair system and metadata that can inform rights that creators hold.




\section{Limitation}
While we were trying to provide a comprehensive analysis of compensation seeking behaviors on Thingiverse with 3D contents shared for free, there are still several limitations. 
First, 
although we also used the text classification model to discover more potential compensation seeking designs in addition to URL analysis, this does not guarantee that we discovered `all' possible cases on Thingiverse, especially for creators who does not directly seek compensation in the description field (e.g., one creator only left his personal website in description where the link to donation account is
on his/her personal website). 
Second, in regard to the 'tip designer', we collected all things made by creators who enabled them as compensation seeking designs. 
However, some may not have a strong intention to seek compensation but just turned on the function as there were people who wanted to leave a tip as we discovered the case from the survey: \textit{``I only have one design available [...] I only enabled the `Tip Designer' option after users requested it to show their appreciation.''}
Third, not all creators classified their designs correctly on Thingiverse, which could potentially influence the analysis that are based on category information. Some examples include models that are either overlap (such as `Models' and `Toys \& Games') or can be easily misunderstood by creators who chose the categories, such as the '3D Printing' category. From the subcategories, it can be guessed that this is mostly for designs that are related to 3D printers (e.g., printer parts, accessories, and testing models), but there were some users misunderstood this as 3D 'printable' models in general. Thus a potential future work is to come up with an efficient and accurate way of verifying the categorization of things.

There also exist limitations in our survey design and analysis. As there were no compensation promised for participation, responses collected could be potentially biased to the users who share designs for altruistic reasons, expecting no returns. 

Also, some findings are drawn from the empirical evidences. The analysis of compensation behavior could be more well-support with more detailed information and data, such as statistics of compensations received by creators and how the creators' overall incomes are influenced by the compensation (e.g. does compensation actually increase creator's household income? Is compensation a stable and sufficient way to sustain home economy?).\par








\section{Conclusion}
This work discusses the potential opportunities in online 3D printing communities for designers to get additional compensation. We analyze the type of compensation methods and behaviors, the type of content that attracts compensation, additional benefits and rewards for compensation, and how such trend increases over time. We find that there exists various opportunities in 3D printing communities, such as providing additional content, professional consultation, printing service, that could help designers get additional compensation. We also discuss the motivation and expectation of designers' compensation in these communities, specifically on how the revenues they received influence their will to continue producing creative works. Although there is a diverse gap between ideal compensation and the current status, it still shows a wide variety of socio-economic opportunities for designers. We conclude by discussing emergent issues, concerns, and implications for potential opportunities and jobs for community members in the future.

\bibliographystyle{ACM-Reference-Format}
\bibliography{main}

\end{document}